\documentclass[aps,prd,groupedaddress,showpacs,showkeys]{revtex4}
\usepackage{epsfig}
\newcommand{\eq}{\begin{eqnarray}}
\newcommand{\en}{\end{eqnarray}}
\newcommand{\bea}{\begin{eqnarray}}
\newcommand{\eea}{\end{eqnarray}}

\newcommand{\ba}[1]{\begin{eqnarray} \label{#1}}
\newcommand{\ea}{\end{eqnarray}}
\newcommand{\vprod}[2]{{\mathbf{#1}\!\cdot\!\mathbf{#2}}}

\begin{document}
\title{Lepton Flavor Violating Photoleptonic Effect}
\author{Edson Carquin}
\email{edson.carquin@usm.cl}
\author{Yuri Ivanov\footnote{On leave of absence from Joint Institute for Nuclear
Research, Dubna, Russia}}
\email{yuri.ivanov@usm.cl}
\author{Sergey Kovalenko}
\email{sergey.kovalenko@usm.cl}
\author{Ivan Schmidt}
\email{ivan.schmidt@usm.cl}
 \affiliation{Centro de Estudios Subat\'omicos(CES),
Universidad T\'ecnica Federico Santa Mar\'\i a, \\
Casilla 110-V, Valpara\'\i so, Chile}

\begin{abstract}
We study lepton flavor violating analogs of the photoelectric
effect, with a final $\mu$ or $\tau$ instead of an electron:
$\gamma e\rightarrow \mu$ and $\gamma e\rightarrow \tau$. On the basis
of the general parametrization of the matrix element of the
electromagnetic current we estimate the upper limits for the cross
sections and event rates of these processes, imposed by the current
experimental bounds on $\mu\rightarrow e \gamma$ and
$\tau\rightarrow e \gamma$ decays.
\end{abstract}
\pacs{PACS numbers:} \maketitle

\section{Introduction}

Lepton Flavor Violation (LFV) has become an experimental fact after the observation of neutrino oscillations, and
this leads immediately to the conclusion of the existence of LFV processes in the sector of charged leptons.
However, the amount of LFV transmitted from the neutrino sector to the sector of charged leptons by Standard Model (SM) loops
is extremely small, leaving no chance for their experimental observation. On the other hand interactions beyond the SM can potentially
induce LFV directly in the sector of charged leptons. Therefore, any observation of an LFV transition of charged leptons would
be a signal of physics beyond the SM. This is the motivation for the theoretical and experimental studies of LFV processes with
charged leptons. Among them the most attention has been paid to muon-electron nuclear conversion, muon(electron)-nucleon scattering
as well as to LFV in decays of mesons, muon, tau (for reviews see, for instance, Refs. \cite{LFV-revs}). It was recognized that different processes may have
quite different sensitivity to the LFV and can shed light on complementary aspects of the underlying physics being, in general, dependent
on different combinations of fundamental parameters of new physics.
Therefore, looking for new processes potentially capable to render new information on the origen of LFV represents an important quest both for theory and experiment.

In the present work we examine a new class of the LFV processes in the charged lepton sector, induced by real photon beams.
This is the LFV version of the photoelectric effect, with a muon or tau in the final state instead of the usual photoelectron.
Our analysis is based on the general parametrization of the electromagnetic current in terms of LFV form factors, without
any reference to the underlying physics behind the LFV electromagnetic transitions of leptons.
The same form factors describe $\mu\rightarrow e \gamma$ and $\tau\rightarrow e \gamma$ decays and, therefore, they are limited by
the existing results on the experimental searches for these processes. We use these limits to predict upper bounds on the total
cross section and event rate of $\gamma e \rightarrow \mu$ and $\gamma e\rightarrow \tau$ for the initial electrons bound to atoms.

\section{Matrix Element of LFV Photo-Leptonic Effect}

The amplitud for the transition $\gamma l_i \rightarrow l_f$, induced by real photons,
can be written in the following standard form
\ba{ME}
  \tilde{\mathcal{M}}_{fi} =
    \int e_{\mu}\langle l_f | J_{em}^{\mu}(x) | l_i \rangle
    e^{-i{k\cdot x}} d^{4}x,
\ea
where $e_{\lambda}$ is the photon polarization 4-vector.

The most general form of the leptonic matrix element in Eq. (\ref{ME}), consistent
with Lorentz covariance and conservation of electric current, is
\ba{GP}
  \langle l_f | J_{em}^{\mu}(x) | l_i \rangle & = &
    \bar{\psi}_f(x)\left[(f^{fi}_{E0}(\hat{k}^2)
    + \gamma_5 f^{fi}_{M0}(\hat{k}^2))\gamma_{\nu}\left(g^{\mu\nu}
    - \frac{\hat{k}^{\nu} \hat{k}^{\mu}}{\hat{k}^{2}}\right)
    + (f^{fi}_{M1}(\hat{k}^2) + \gamma_5 f^{fi}_{E1}(\hat{k}^2))
      i \sigma^{\mu\nu} \frac{\hat{k}_{\nu}}{m_{e}} \right] \psi_{i}(x)\\
\nonumber
  & \equiv & \bar{\psi}_f(x) \Gamma^{\mu}_{fi}(\hat{k}) \psi_{i}(x),
\ea
where $\psi_{i(f)}$ are the wave functions of the initial(final) lepton. We defined
$\hat{k}_{\mu} = i (\overleftarrow{\partial}_{\mu} + \overrightarrow{\partial}_{\mu})$
the differential operator of 4-momentum transfer, with the first derivative acting to
the left hand side and the second one to the right hand side. For the case of free
leptons this operator is to be replaced as
$\hat{k}_{\mu} \rightarrow k_{\mu} = p_{(f)\mu} - p_{(i)\mu}$,
where $p_{(i)\mu}$ and $p_{(f)\mu}$ are 4-momenta of the initial and final leptons
respectively. For convenience we also introduced the function $\Gamma^{\mu}_{fi}$.
In Eq. (\ref{GP}) the functions $f_{E0}(k^2), f_{M0}(k^2)$ and $f_{E1}(k^2), f_{M1}(k^2)$
are the conventional monopole and dipole electric and magnetic transition form factors.
From T-invariance it follows that all the above form factors are real and symmetric
\ba{T-inv}
  f^{if}_{E0} = f^{fi}_{E0}, \ \ \  f^{if}_{M0} = f^{fi}_{M0},\ \ \
  f^{if}_{E1} = f^{fi}_{E1}, \ \ \  f^{if}_{M1} = f^{fi}_{M1}.
\ea
Thus, the same set of form factors describe $\gamma l_i \rightarrow l_f$ and
$l_i\rightarrow l_f \gamma$ processes. The monopole form factors must satisfy
the finiteness conditions
\ba{finit}
  f^{if}_{E0}(0)= f^{if}_{M0}(0)= 0
\ea
and, therefore, they do not contribute to the $\gamma l_i \rightarrow l_f$ processes
with a real photon, which has $k^2=0$.

Substituting the expression (\ref{GP}) into Eq. (\ref{ME}) and integrating by parts
we obtain
\ba{ME-1}
  \tilde{\mathcal{M}}_{fi} = \int \bar{\psi}_f(x) e_{\mu} \Gamma^{\mu}_{fi}(k)
                             \psi_{i}(x), e^{-i{k\cdot x}} d^{4}x
\ea
where the vertex function $\Gamma^{\mu}$ is the function defined in Eq. (\ref{GP}).

Let us turn to the LFV photoeffect: $\gamma e \rightarrow l$ with $l = \mu, \tau$.
In this case the initial lepton is the electron bound to the atom with energy
$\varepsilon_e = m_e-I$, where $I$ is the corresponding value of the ionization energy.
The incident real photon with energy $\omega$ and momentum $\mathbf{k}$ hits
the atomic electron and creates the final lepton $l$ with energy $\varepsilon_l$
and momentum $\mathbf{p}_l$. Therefore, we can rewrite Eq. (\ref{ME}) in the 3-dimensional
transversal gauge, $e_{\mu}k^{\mu} = 0$ with $e_{\mu}=(0, \mathbf{e})$, in the form
\ba{ME-I}
  \tilde{\mathcal{M}}_{l e} = 2 \pi \delta(\varepsilon_i+\omega-\varepsilon_l)
    \int \bar{\psi}_{l}(\mathbf{x}) \
    (\mathbf{e\cdot \Gamma})_l\  \psi_{e}(\mathbf{x})
    e^{i\mathbf{k x}} d^{3}x
  \equiv 2 \pi \delta(\varepsilon_i+\omega-\varepsilon_l) \mathcal{M}_{le}\ ,
\ea
where $\psi_{e,l}(\mathbf{x})$ are the spacial wave functions of the initial
electron and final lepton. Here we also introduced the reduced matrix element
$\mathcal{M}_{le}$ of $\gamma e\rightarrow l$ transition. In virtue of Eq.~(\ref{finit})
the product of vertex function and the photon polarization vector is given by
\ba{Vert} (\mathbf{e\cdot \Gamma})_l = \frac{i}{m_{e}}
  (f_{M1}^{el} + \gamma_5 f_{E1}^{el})
  (\mbox{\boldmath $\gamma\cdot$}\mathbf{e})
  (\gamma^0\omega - \mbox{\boldmath $\gamma\cdot$}\mathbf{k})
\ea
Here, $f_{E1}\equiv f_{E1}(0),\ f_{M1}\equiv f_{M1}(0)$ and $\omega=|\mathbf{k}|$
is the photon energy. In what follows we consider the LFV photoeffect from the
ground state atomic level of a hydrogen-like ion with atomic number $Z\ll 137$.
The latter condition allows one to derive the final result for the cross section
in an explicit analytic form. The non-relativistic ground state electron wave
function is
\ba{gs-WF}
  \psi_{0}(r) = \frac{(Z e^2 m_e)^{3/2}}{\sqrt{\pi}} e^{-Z e^2 m_e r}.
\ea
As is known \cite{Lifshitz}, despite the fact that the initial electron is non-relativistic,
for a self-consistent treatment of the photoeffect it is necessary to take
into account relativistic corrections to its wave function at least to first order in
the small parameter $Z e^2 \ll 1$.

The initial electron wave function corrected in this way is given by
\cite{Lifshitz}
\ba{gs-RWF}
  \psi_e = \left(1 - \frac{i}{2 m_e} \gamma^0\mathbf{\gamma \nabla}\right)
           \frac{u_e}{\sqrt{2 m_e}} \psi_0,
\ea
where $u_e$ is bispinor amplitude of the electron in the rest frame, normalized by
$\bar{u_e}u_e=2 m_e$.

We write the wave function of the final lepton in the form
\ba{fl-RWF}
  \psi_{l} & = & \frac{1}{\sqrt{2\varepsilon_l}}
  \left(u_{l}e^{i\mathbf{p}_l\cdot\mathbf{r}}+\psi^{(\!1)} \right),
\ea
where the term $\psi^{(\!1)}$ represents the leading $Z e^2$ Coulomb correction.
Its Fourier transform is \cite{Lifshitz}
\ba{RWF-Coulomb}
  \bar{\psi}^{(\!1)}_{-\mathbf{k}}
    = \int d^3 x \bar{\psi}^{(\!1)}(\mathbf{x}) e^{i \mathbf{kx}}
    = 4 \pi Z e^{2}\bar{u}_l\frac{2\varepsilon_l
      \gamma^{0} + \mbox{\boldmath $\gamma\cdot$}
      (\mathbf{k}-\mathbf{p}_l)}{(\mathbf{k}^{2}-\mathbf{p}_l^{2})
      (\mathbf{k}-\mathbf{p}_l)^{2}}
\gamma^{0}
\ea

Now, substituting Eqs. (\ref{gs-WF})-(\ref{RWF-Coulomb}) into Eq. (\ref{ME-I}),
we obtain to first order of perturbation theory in $Z e^2$, the following
expression for the reduced matrix element
\ba{ME-expression}
  \mathcal{M}_{le} =
    \frac{4 \pi^{1/2}(Ze^{2}m_e)^{5/2}}{(\varepsilon_l \ m_e)^{1/2}
    (\mathbf{k}-\mathbf{p}_l)^2} \bar{u}_l A_l u_e
\ea
with
\ba{Def-A}
  A_l = a (\mbox{\boldmath $\Gamma\cdot$}\mathbf{e})_l
        + (\mbox{\boldmath $\Gamma\cdot$}\mathbf{e})_l \gamma^0
          (\mbox{\boldmath $\gamma\cdot$}\mathbf{b})
        + (\mbox{\boldmath $\gamma\cdot$}\mathbf{c}) \gamma^0
          (\mbox{\boldmath $\Gamma\cdot$}\mathbf{e})_l
\ea
where
\ba{def-II}
  a = \frac{1}{(\mathbf{p}_l-\mathbf{k})^2}
    + \frac{\varepsilon_l}{m_e}\frac{1}{\mathbf{k}^{2}-\mathbf{p}_l^{2}},~~
  \mathbf{b} = \frac{\mathbf{p}_l-\mathbf{k}}{2m_e(\mathbf{p}_l-\mathbf{k})^{2}},~~
  \mathbf{c} = \frac{\mathbf{k}-\mathbf{p}_l}{2m_e(\mathbf{k}^{2}-\mathbf{p}_l^{2})}
\ea

The differential cross section summed over the final lepton polarization and averaged
over the initial electron one takes the form
\ba{cs}
  d\sigma(\gamma e\rightarrow l) =
    \frac{4 \alpha_{em}^6 Z^5 m_e^5}{\omega (\mathbf{k}-\mathbf{p}_l)^4} |\mathbf{p}_l|
    \mbox{Tr} \left[(\gamma^0\varepsilon_l-\mbox{\boldmath$\gamma$}\mathbf{p}_l + m_l) A_l
    (\gamma^0+1) \gamma^0 A_l^{\dagger}\gamma^0\right] d\Omega~.
\ea
Carrying out the trace one can obtain the following expression
\ba{Tr}
  \mbox{Tr}[...] &=& \frac{8}{m_l^2}
    \left\{
      \left[ |f_{E1}^{el}|^2 + |f_{M1}^{el}|^2 \right] T_+
    + \left[|f_{E1}^{el}|^2 - |f_{M1}^{el}|^2 \right] T_-
    \right\}~,
\ea
where
\ba{TrP}
  T_+&=& 2 \left[
             \vprod{k}{d}~\vprod{p}{c}
           - \vprod{k}{p}~\vprod{c}{d}
           + \varepsilon_l\left(\vprod{k}{b}~\vprod{k}{c}-\omega^2\,\vprod{b}{c}\right)
           \right]
         + \left(\varepsilon_l-\vprod{k}{p}/\omega\right) \mathbf{d}^2~,\\
  \label{TrM}
  T_-&=& 2\,m_l
           \left[\vprod{k}{b}~\vprod{k}{c}
         + \omega^2\left(2~\vprod{b}{e}~\vprod{c}{e}-\vprod{b}{c}\right)
           \right]
\ea
and $\mathbf{d}=a\mathbf{k}-\omega(\mathbf{b}+\mathbf{c})$. In the total cross
section term $T_-$ drops out since $T_- \sim p_y^2-p_x^2$ and, therefore, integration over
the angles gives zero.
The final result for the
total cross section can be written in terms of dimensionless
variables $t = \varepsilon_l/m_l$, $u = \omega/m_l$ and $v =
\sqrt{t^2-1}$ in the following form
\ba{total_cross}
  \sigma(\gamma e\rightarrow l) &=&
    16\, \alpha_{em}^6 Z^5 \,
    \frac{m_e^5}{m_l^7}\,
    \left[\left(f_{E1}^{el}\right)^2 + \left(f_{M1}^{el}\right)^2 \right]\,
    \frac{v}{u}\, F(t,u,v)
  ~,
\ea
where
\ba{total_cross_F}
F(t,u,v) &=&
   \frac{P(t,u,v)             }{3     (u^2-v^2)^6}
 + \frac{t (u^2+v^2) - 2 u v^2}{2 u v (u^2-v^2)^2}\,\log\left(\frac{u-v}{u+v}\right)
~, \\
P(t,u,v) &=&
     3 t u^6     \left(8 (t+1)^2+u^2\right)
  -  2   u^4 v^2 \left(24 t^3-4 t (10-u (4+3 u))+u (32+u (16+3 u))\right) + \\ \nonumber
 &+& 2   u^2 v^4 \left(12 t (t\!-\!1)^2\!+\!16 u (t\!-\!2)\!+\!21 t u^2\!+\!3 u^3\right)
  + 2   u^2 v^6 \left(16\!-\!12 t\!+\!3 u\right)
  + 3       v^8 \left(t\!-\!2 u\right)
~.
\ea
Taking into account that for both the muon and $\tau$-lepton $\varepsilon_e/m_l \ll 1$,
the variable $t$ can be approximated as $t=u+\varepsilon_e/m_l \approx u$ and the
above expressions can be written (with accuracy of $\varepsilon_e/m_l$) as follows
\ba{total_cross_F_approx}
F(t,u,v) &\approx&
  \frac{u}{(u^2-v^2)^2}\,
  \left[
       \frac{64 u^4 + 80 u^3 -10 u^2 -32 u - 3}{3(u^2-v^2)^4}
     + \frac{1}{2 u v}\,\log\left(\frac{u-v}{u+v}\right)
  \right]~.
\ea
%
%
From this expression it can be seen that the total cross sections of both LFV processes steeply rise with the photon energy, in contrast
to the case of the classic photoelectric effect with an electron in the final state, whose cross section decreases with the photon energy.
This feature is manifest in Fig. 1, to be discussed in the next section.

\section{Experimental constraints on the Form Factors}

Since the same form factors $f^{if}_{E1,M1}$ determine both $\gamma l_i \rightarrow l_f$ and
$l_f \rightarrow l_i \gamma$ processes, we can derive upper limits on $f^{e\mu}_{E1,M1}$ and
$f^{e\tau}_{E1,M1}$ from the existing experimental bounds on $\mu \rightarrow e \gamma$ and
$\tau \rightarrow e \gamma$ \cite{BABAR,SINDRUM,MEGA}
\ba{exp-1}
  Br(\tau^- \rightarrow e^- \gamma) &=&
    \frac{\Gamma(\tau^-\rightarrow e^{-}\gamma)}{\Gamma_{\tau}} \leq 1.1 \times 10^{-7},\\
\label{exp-2}
  Br(\mu^-  \rightarrow e^- \gamma) &=&
    \frac{\Gamma(\mu^-\rightarrow e^{-}\gamma)}{\Gamma_{\mu}} \leq 1.2 \times 10^{-11}
\ea
and then apply these limits for the evaluation of upper bounds on the processes in which we are interested:
$ \gamma e \rightarrow \mu$ and $ \gamma e \rightarrow \tau$. In Eqs.~(\ref{exp-1}), (\ref{exp-2}) we use
$\Gamma_{\tau} = 2.26\times 10^{-5} $ MeV and $\Gamma_{\mu} = 3\times 10^{-16}$ MeV for
total decay widths of the $\tau$ and $\mu$.

The decay rates are given by
\ba{Rate}
  \Gamma(l^-\rightarrow e^{-}\gamma) &=&
    \frac{m_l^3}{8 \pi m_e^2}\left(|f^{el}_{E1}|^2 + |f^{el}_{M1}|^2\right),
\ea
where $l=\mu, \tau$.

Comparing Eqs. (\ref{Rate}) with Eqs. (\ref{exp-1}) and (\ref{exp-2}) we get upper limits on the absolute
values of the dipole form factors
\ba{lim}
  |f^{e\mu}_{E1}|^2 + |f^{e\mu}_{M1}|^2 \leq 2.0\times 10^{-32},\ \ \ \ |f^{e\tau}_{E1}|^2 + |f^{e\tau}_{M1}|^2 \leq 3.0\times 10^{-21}.
\ea
Substituting these limits in Eqs. (\ref{total_cross}) we evaluate the upper limits
on the total cross sections of the photomuonic and phototauonic effects. An exclusion plot for the case
of a lead (Pb) atom is shown in Fig. 1.
Thus, the upper limits on the form factors extracted from the experimental data on
$\mu \rightarrow e \gamma$ and $\tau \rightarrow e \gamma$ decays impose very strong limits on
their inverse processes $\gamma e\rightarrow \mu(\tau)$ with atomic electrons.

In order to assess if such small cross sections leave any chance for the experimental observation of
the LFV Photo-leptonic processes under question, we estimate
the corresponding reaction rate
\ba{ra}
R=\sigma(\gamma e\rightarrow l)\cdot L
\ea
where $L$ is the target luminosity $L$ of the incident
photon beam. Taking roughly that all the incident photons
are absorbed within a target depth equal to the photon conversion
length $\kappa$  we estimate the target luminosity as
\ba{Lumin}
L = Z\ \frac{\kappa}{A}\  {\cal F}_{\gamma}\times 4.35\times 10^{-16}{\rm fb^{-1}/s},
\ea
where $A$, $Z$ and $\kappa$ are target material
nuclear mass number in atomic units, atomic number and the conversion length in $\rm{g\cdot cm^{-2}}$, respectively.
The photon flux ${\cal F}_{\gamma}$  is measured in $s^{-1}$.

As an example we consider a lead (Pb) target
with $A=207.2$, $Z=82$ and $\kappa= 7.46\ \rm{g\cdot cm^{-2}}$. Its corresponding luminosity is
\ba{Lumin-Pb}
L_{Pb} = 1.3 \times 10^{-15}\cdot {\cal F}_{\gamma}{\rm fb^{-1}/s},
\ea
With this luminosity we estimate the number of the LFV events
\ba{year}
R(\gamma e\rightarrow \mu) &\approx& 2.0\times 10^{-40}\cdot {\cal N}_{\gamma}\ \ \ \ \mbox{for} \ \ \ \ \omega=1 {\rm GeV}\\
R(\gamma e\rightarrow \tau)&\approx& 1.0\times 10^{-40}\cdot {\cal N}_{\gamma} \ \ \ \ \mbox{for} \ \ \ \ \omega=5 {\rm GeV}\\
\ea
where ${\cal N}_{\gamma}$ is the number of the photons absorbed in the lead target. This result means that the observation of one LFV event
would require a photon energy deposit to the target of about $10^{30}$J. It is clear that these conditions are unrealistic. Higher event rates
correspond to photon energies too high to be achieved in near future experiments with beams of sufficiently high intensity.
Thus we conclude that the LFV processes $\gamma e\rightarrow \tau$ and $\gamma e\rightarrow \tau$ are experimentally unobservable
under the existing experimental limits on $\mu \rightarrow e \gamma$ and $\tau \rightarrow e \gamma$.
In the other words the latter processes are much more sensitive to the LFV than the ones studied in the present paper.

\section{Discussion and Conclusions}

We have studied LFV photoproduction of $\mu$ and $\tau$ on atomic electrons. We extended the conventional formalism, for
ordinary photoelectric effect with a final electron, to the case of $\gamma e\rightarrow \mu$ and $\gamma e\rightarrow \tau$ processes.
We have provided a general parametrization of the operator of electromagnetic current, instead of the commonly used parametrization
of its matrix elements. This representation allowed us to consistently treat the off-mass-shell initial atomic electron in terms of LFV analogs
of the conventional monopole $f_{M0}, f_{E0}$ and dipole $f_{M1}, f_{E1}$ electromagnetic form factors of the electron. The studied LFV processes
with real photons are independent of the monopole form factors, depending only on the dipole LFV form factors $f^{e\mu}_{M1,E1}, f^{e\tau}_{M1,E1}$.
These form factors are also involved in $\mu \rightarrow e \gamma$ and $\tau \rightarrow e \gamma$ decays, whose rates are limited
by the existing experimental data. Using these experimental limits we extracted upper bounds on the dipole LFV form factors and predicted
the total cross sections of $\gamma e\rightarrow \mu$ and $\gamma e\rightarrow \tau$ processes. We also evaluated prospects for their experimental observation
and arrived at a result that the event rate leaves no chance for this observation in any realistic experiment.
In other words, the experiments looking for $\mu \rightarrow e \gamma$ and $\tau \rightarrow e \gamma$
decays are much more sensitive to LFV than the above studied photoproduction processes $\gamma e\rightarrow \mu$ and $\gamma e\rightarrow \tau$.

\begin{acknowledgments}

This work was supported by CONICYT (Chile) under grant PBCT/No.285/2006.

\end{acknowledgments}

\newpage
\begin{figure}[htb]
\centerline{
  \scalebox{0.65}{\includegraphics{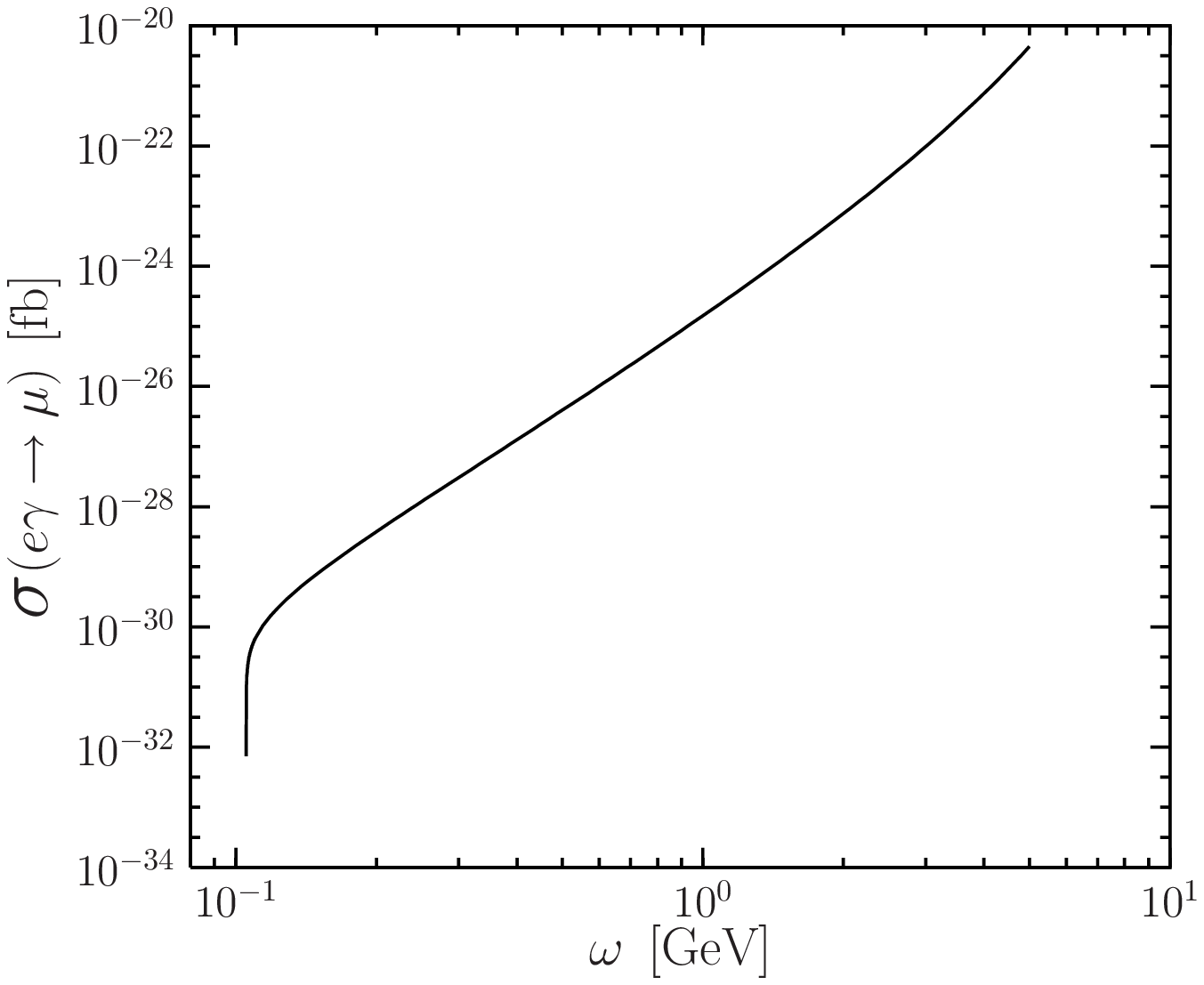}}~~~
  \scalebox{0.65}{\includegraphics{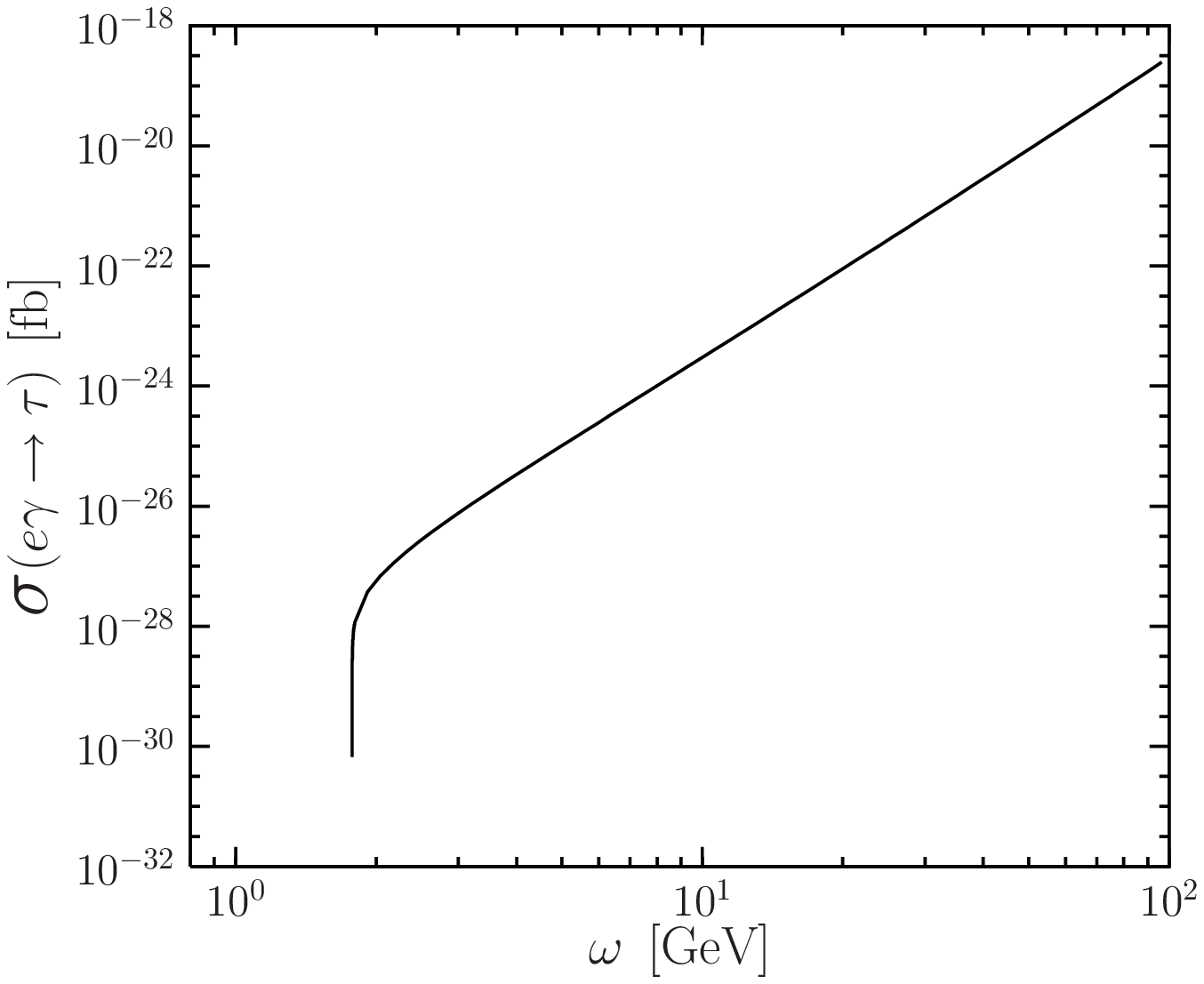}}
}
\caption{The total cross sections of the process $\gamma e\rightarrow \mu,\tau$ with the lead $Z=82$ atomic electron. The regions above the curves are excluded
by the present experimental limits on $\mu,\tau\rightarrow e \gamma$ decays.
}
\end{figure}


\begin{thebibliography}{3}

\bibitem{LFV-revs} J.D. Vergados, Phys. Rep. {\bf 133}, 1 (1986); T.S. Kosmas, G.K. Leontaris, and J.D. Vergados,
Prog. Part. Nucl. Phys. {\bf 33}, 397 (1994); W.J. Marciano, "Lepton flavor violation, summary and
 perspectives",  Honolulu-Hawai, USA, October 2-6, 2000, http://meco.ps.uci.edu/lepton\_workshop.
%
\bibitem{Lifshitz} E.M. Lifshitz et.al., \textit{Quantum Electrodynamics,
Volume 4 of Course of Theoretical Physics, Second Edition,}
Izdatel'stvo 'Nauka', Moscow 1980.
%
\bibitem{BABAR} B. Aubert et al. [BABAR Collaboration], arXiv:hep-ex/0508012.
%
\bibitem{SINDRUM} U. Bellgardt et al. [SINDRUM Collaboration], Nucl. Phys. B 299, 1 (1988).
%
\bibitem{MEGA} M. L. Brooks et al. [MEGA Collaboration], Phys. Rev. Lett. 83 (1999) 1521.
%
\end{thebibliography}
\end{document}